\documentclass[acmtog]{acmart}
\acmSubmissionID{290}

\usepackage{booktabs,enumitem} 
\usepackage{amsmath,amsthm,units}

\newcommand{\R}{\mathbb R}
\newcommand{\changed}[1]{#1}

\renewcommand{\epsilon}{\varepsilon}
\renewcommand{\L}{\mathcal L}

\usepackage[utf8]{inputenc}

\let\oldstar\star
\renewcommand{\star}{{\oldstar}}

\usepackage{gensymb}
\usepackage[ruled]{algorithm2e} 
\usepackage{microtype}
\usepackage{wrapfig}

\SetAlFnt{\small}
\SetAlCapFnt{\small}
\SetAlCapNameFnt{\small}
\SetAlCapHSkip{0pt}

\usepackage{pdfpages}

\setlist[itemize]{leftmargin=*}

\setcopyright{rightsretained} 
\acmJournal{TOG}
\acmYear{2021}\acmVolume{40}\acmNumber{4}\acmArticle{166}\acmMonth{8} \acmDOI{10.1145/3450626.3459797}

\begin{document}
\title{HodgeNet: Learning Spectral Geometry on Triangle Meshes}

\author{Dmitriy Smirnov}
\email{smirnov@mit.edu}
\author{Justin Solomon}
\email{jsolomon@mit.edu}
\affiliation{%
  \institution{Massachusetts Institute of Technology}
  \streetaddress{77 Massachusetts Avenue}
  \city{Cambridge}
  \state{MA}
  \postcode{02139}
  \country{USA}}

\begin{abstract}
Constrained by the limitations of learning toolkits engineered for other applications, such as those in image processing, many mesh-based learning algorithms employ data flows that would be atypical from the perspective of conventional geometry processing.  As an alternative, we present a technique for learning from meshes built from standard geometry processing modules and operations.  We show that low-order eigenvalue/eigenvector computation from operators parameterized using discrete exterior calculus is amenable to efficient approximate backpropagation, yielding spectral per-element or per-mesh features with similar formulas to classical descriptors like the heat/wave kernel signatures.  Our model uses few parameters, generalizes to high-resolution meshes, and exhibits performance and time complexity on par with past work.
\end{abstract}



\begin{teaserfigure}
    \centering
    \includegraphics[width=\linewidth]{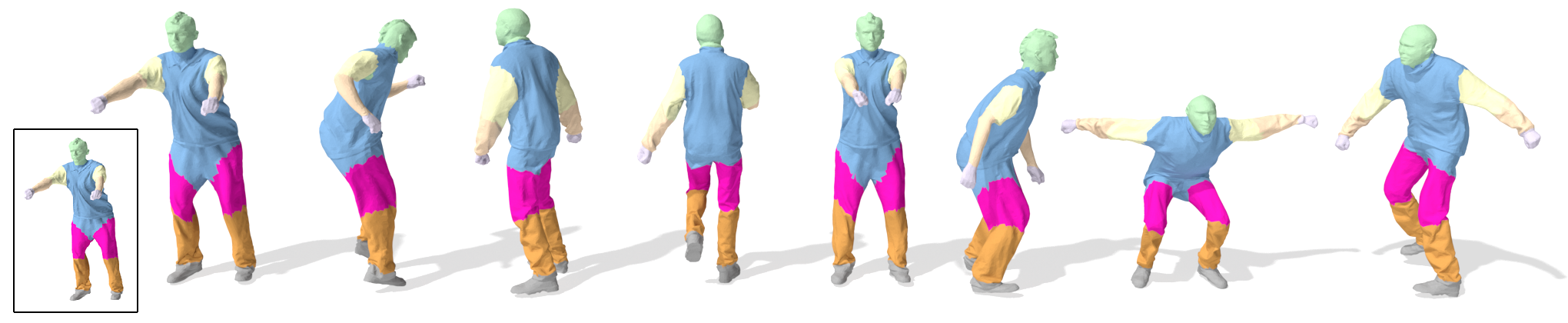}
    \caption{Mesh segmentation results on the full-resolution MIT animation dataset. Each mesh in the dataset contains 20,000 faces (10,000 vertices). We show an example ground truth segmentation in the bottom-left. In contrast to previous works, which downsample each mesh by more than $10\times$, we efficiently process dense meshes both at train and test time.}
    \label{fig:humans_highres}
\end{teaserfigure}

\maketitle

\section{Introduction}

Data-driven algorithms have altered the way we approach problem-solving in computer graphics.  Machine learning tools garner top performance for tasks like image editing, user interaction, image synthesis, and layout, supported by large, well-curated datasets.  Yet, while learning tools for areas like computational photography and rendering are widely adopted, another branch of graphics has been resistant to change:  mesh-based shape analysis.

Numerous technical challenges preclude modern learning methods from being adopted for meshes.  Deep learning---arguably the most popular recent learning methodology---relies on \emph{regularity} of the data and \emph{differentiability} of the objective function for efficiency.  For example, convolutional neural network (CNN) training is built on high-throughput processing of images through convolution and per-pixel computations to obtain gradients with respect to network weights, required for stochastic gradient descent.

Meshes, a primary means of representing geometry in graphics, defy the considerations above.  They come as sparse, irregular networks of vertices varying in number; the same piece of geometry easily can be represented by multiple meshes and at multiple resolutions/densities.  Advances in graph neural networks (GNNs) have as a byproduct helped advance mesh processing, but typical graphs in geometry processing are fundamentally different from those in network science---vertices have low valence, are related through long chains of edges, can be connected in many roughly-equivalent ways, and can be deformed through rigid motions and isometries.  

The end result is that mesh-based learning architectures often contort input data to make it compatible with existing learning toolkits.  Restricting to GPU-parallel, regularly-structured computation is a vast limitation for mesh analysis.  For example, while geometry processing algorithms frequently rely on inversion and eigenanalysis of sparse matrices, these operations are hardly compatible with deep learning.  Instead, mesh-based learning algorithms differ from successful non-learning geometry processing algorithms, relying on easily differentiated/parallelized local operations.

In this paper, we ask whether we can invert this relationship: Rather than inventing new data streams for geometry processing to suit existing learning algorithms, can we develop learning methodologies from successful geometry processing techniques?

We target applications in shape analysis using a prevalent tool in that domain, spectral geometry.  Myriad shape analysis algorithms follow a similar template, building a positive (semi)definite matrix whose sparsity pattern is inherited from the mesh and then using its spectral structure to infer information about meshed geometry.  Some examples include the Laplacian operator, the bilaplacian operator, the Dirac operator, and modal analysis.  Our broad goal---also explored in some past work (see \S\ref{sec:related})---is to \emph{learn} the entries of this operator as functions of local geometry.

Unlike past work, however, we observe that classical shape analysis relies on \emph{near-zero eigenvalues} of these operators (and the corresponding eigenvectors); high-frequency information is discarded.  This is a key reason why classical geometry processing algorithms involve sparse matrix inversion and partial computation of eigenvalues.  Partial eigenvector computation from a sparse matrix, however, is incompatible with most existing learning pipelines, so learning algorithms that use low-order eigenanalysis typically precompute the relevant eigenvectors from a fixed operator.  Approaches to operator learning work with operator-vector products (rather than inverting the operator), restrict to a pre-computed basis, or compute the full spectrum as a dense matrix, which is prohibitive for large meshes.

In this paper, we approximately differentiate through sparse operator construction for one class of operators motivated by discrete differential geometry.  As a result, we can learn operators whose entries are functions of local geometry, which together modify the spectral structure of the operator---a global computation.  Our method is competitive with existing mesh-based learning tools while being implemented from standard components of the geometry processing toolkit, and we show how to handle boundary conditions and vector-valued data.  

We make some unconventional design decisions that resemble geometry processing rather than deep learning.  For instance, our spectral computation and operator construction are implemented using sparse linear algebra on the CPU, 
and we implement geometry processing-specific strategies for data augmentation that promote resolution independence.  These decisions do not hamper efficiency of our method relative to past work.

\paragraph*{Contributions.} We present a lightweight model for learning from triangle meshes, with or without boundary.  Contributions include:
\begin{itemize}
    \item a learnable class of sparse operators on meshes built from standard constructions in discrete exterior calculus;
    \item parallelizable algorithms for differentiating eigencomputation from these operators, including approximate backpropagation without sparse computation;
    \item end-to-end architectures for learning per-element or per-mesh features starting from mesh geometry without additional features; 
    \item simple strategies for data augmentation and other practical techniques to improve performance of our method; and
    \item experiments demonstrating effectiveness in shape analysis tasks, including \changed{the generalization of our model} to high-resolution meshes that are too dense to be compatible with related methods.
\end{itemize}

\section{Related Work}\label{sec:related}

Machine learning from geometry is becoming a popular subfield of graphics and vision.  \citet{bronstein2017geometric} provide a broad overview of challenges in this discipline; here, we focus on work directly related to our task of learning from meshed geometry.

\subsection{Spectral Shape Analysis}

Our method is built on ideas from spectral geometry, which captures shape properties through the lens of spectral (eigenvalue/eigenvector) problems.  \citet{wang2019intrinsic} provide a comprehensive introduction to this approach to geometry processing.

The \emph{Laplace--Beltrami} (or, \emph{Laplacian}) operator is ubiquitous in  spectral geometry processing.  Most relevant to our work, numerous per-vertex and per-mesh features have been built from Laplacian eigenvalues and eigenvectors, including the global point signature \citep{rustamov2007laplace}, the heat kernel signature \citep{sun2009concise}, the wave kernel signature \citep{aubry2011wave}, and the heat kernel map \citep{ovsjanikov2010one}.  These descriptors underlie algorithms for tasks as varied as symmetry detection \citep{ovsjanikov2008global}, correspondence \citep{ovsjanikov2012functional}, shape recognition \citep{reuter2006laplace,bronstein2010shape}, and shape retrieval \citep{bronstein2011shape}---among countless others.

The Laplacian is popular given its multiscale sensitivity to intrinsic geometry, but recent work proposes replacements sensitive to other aspects of geometry like extrinsic deformation.  Examples include the Dirac operator \citep{liu2017dirac,ye2018unified}, modal analysis \citep{hildebrandt2012modal,huang2009shape}, the Hamiltonian \citep{choukroun2018sparse}, the curvature Laplacian \citep{liu2007mesh}, the concavity-aware Laplacian \citep{au2011mesh,wang2014spectral}, the volumetric Laplacian \citep{raviv2010volumetric}, and the Dirichlet-to-Neumann operator \citep{wang2018steklov}.  Other works add invariances to the Laplacian, e.g., to local scaling \citep{bronstein2010scale} or affine deformation \citep{raviv2011affine}, while others incorporate local features like photometric information \citep{kovnatsky2011photometric,spagnuolo2012affine}.  Nearly all these algorithms---with the notable exception of volumetric methods \citep{raviv2010volumetric,wang2018steklov}---follow the same outline:  Build an operator matrix whose sparsity pattern is inherited from the edges of a triangle mesh and construct features from its eigenvectors and eigenvalues; a widely-used strategy of \emph{truncation} approximates spectral features using partial eigeninformation, usually the eigenvalues closest to $0$.

Other spectral methods use or produce \emph{vectorial} data, working with operators that manipulate tangential fields.  Vector diffusion operators move information along a manifold or surface while accounting for parallel transport \citep{singer2012vector,sharp2019vector}.  The Killing operator also has been applied to intrinsic symmetry detection \citep{ben2010discrete}, segmentation \cite{solomon2011discovery}, deformation \cite{solomon2011killing,claici2017isometry}, level set tracking \citep{tao2016near}, and registration/reconstruction \citep{chan2013reconstruction,slavcheva2017killingfusion}. These methods again analyze a sparse operator built from local features and mesh structure, although there is less agreement on the discretization of operators acting on vector-valued data \citep{de2016vector}.

Spectral representations of geometry can be ``complete'' in the sense that a shape's intrinsic structure or embedding can be reconstructed from the eigenvalues and eigenvectors of certain operators.  For example, the discrete Laplacian determines mesh edge lengths \cite{zeng2012discrete}, and a modified operator adds the extrinsic information needed to obtain an embedding \cite{corman2017functional}.  \citep{boscaini2015shape,corman2017functional,cosmo2019isospectralization} solve related inverse problems in practice.

Transitioning to the next section, an early machine learning method by \citet{litman2013learning} uses regression to learn spectral descriptors on meshes through learnable functions of Laplacian eigenvalues. This method does not learn the operator but rather the way per-vertex features are constructed from Laplacian eigenvalues. \citet{henaff2015deep} propose a similar approach on graphs.

We attempt to generalize many of the methods above.  Rather than defining a ``bespoke'' operator and, mapping from eigeninformation to features for each new task, however, we learn an operator from data.

\subsection{Neural Networks on Meshes}

Many papers propose algorithms for learning from meshes and other geometric representations.  Here, we summarize past approaches for learning features from meshes, although specialized methods for mesh-based learning appear in tasks like generative modeling \citep{liu2020neural,hertz2020deep}, meshing \cite{sharp2020ptn}, and reconstruction \citep{gao2020learning,hanocka2020point}.

\paragraph*{Learning from graphs.}  

Since triangle meshes are structured graphs, algorithms for learning from graphs inspired approaches to learning from meshes.  Indeed, graph neural networks (GNNs) \citep{kipf2017semi} are often used as baselines for geometric learning.

The graph analog of spectral geometry employs Laplacian matrices that act on per-vertex functions.  Graph Laplacians provide a linear model for aggregating information between neighboring vertices. Spectral networks \citep{bruna2013spectral} project per-vertex features onto a low-frequency Laplacian eigenbasis before applying a learned linear operator, followed by a per-vertex nonlinearity in the standard basis; convolution on images can be understood as a spectral filter, so these networks generalize image-based convolutional neural networks (CNNs).  Subsequent work accelerated learning and inference from spectral networks, often using matrix functions in lieu of computing a Laplacian eigenbasis, e.g., via Chebyshev polynomials \citep{defferrard2016convolutional}, random walks \citep{atwood2016diffusion}, or rational functions \citep{levie2018cayleynets}.

\paragraph*{Spatial domain.} 

Many mesh-based learning methods operate in the ``spatial domain,'' relating vertices to their neighbors through constructions like local parameterization or tangent plane approximation.  These methods often can be understood as GNNs with geometrically-motivated edge weights.

Starting with \citep{masci2015geodesic}, many methods define convolution-like operations within local neighborhoods by parameterizing vertices and their $k$-ring neighborhoods.  A challenge is how to orient the convolution kernel, since the tangent plane is different at every point; strategies include taking a maximum over all possible orientations \citep{masci2015geodesic, sun2020zernet}, dynamically computing weights from neighboring features \citep{verma2018feastnet}, aligning to principal curvatures \citep{BoscainiMRB16}, learning pseudo-coordinate functions represented as mixtures of Gaussians \citep{Monti_2017_CVPR}, projecting onto tangent planes \citep{tatarchenko2018tangent}, sorting nearby vertices based on feature similarity \citep{wang20183d}, aligning to a 4-symmetry field \citep{huang2019texturenet}, and weighting by normal vector similarity \citep{song2020meshgraphnet} or directional curvature \citep{he2020curvanet}.  

These and other methods must also define a means of representing localized convolution kernels.  Many choices are available, including localized spectral filters \citep{boscaini2015learning}, B-splines \citep{fey2018splinecnn}, Zernike polynomials \citep{sun2020zernet}, wavelets \citep{schonsheck2018parallel}, and extrinsic Euclidean convolution \citep{schult2020dualconvmesh}.

Additional machinery is needed to compute vectorial features or relate tangent kernels at different vertices---a problem related to choosing a canonical orientation per vertex.  Parallel transport is a choice motivated by differential geometry \citep{pan2018convolutional}, which can be combined with circular harmonics \cite{wiersma2020cnns} or pooling over multiple coordinates \citep{poulenard2018multi} to avoid dependence on a local coordinate system.  \citet{yang2020pfcnn} employ locally flat connections for a similar purpose.

Simple GNN layers like \citep{kipf2017semi} communicate information only among neighboring vertices.  This small receptive field---inherited by several methods above---is a serious challenge for learning from meshes, which are sparse graphs for which a single such layer becomes more and more local as resolution increases.  This issue creates dependency of performance on mesh resolution.

\paragraph*{Mesh-based constructions.} 

While it is valid to interpret meshes as graphs, this neglects the fact that meshes are highly-structured relative to graphs in other disciplines; a few learning algorithms leverage this additional structure to engineer mesh-specific convolutional-style layers.  The popular MeshCNN architecture \citep{hanocka2019meshcnn} learns edge features and performs pooling based on edge collapse operations.  PD-MeshNet \citep{milano2020primal} augments the graph of mesh edges with the graph of dual edges capturing triangle adjacency, with pooling techniques inspired by mesh simplification and dynamic local aggregation using attention.

\paragraph*{Global parameterization.} 

Surface parameterization is a standard technique for texture mapping; some methods parameterize meshes into an image domain on which standard CNNs can be used for learning and inference from pushed-forward features.  \citet{sinha2016deep} pioneered this approach using geometry images \citep{gu2002geometry} for parameterization.  \citet{maron2017convolutional} use seamless toric covers, conformally mapping four copies of a surface into a flat torus; this work was extended by \citet{haim2019surface} to general covers to reduce distortion.  Rendering-based techniques can also be understood as simple parameterizations onto the image plane, e.g., using panoramic \citep{shi2015deeppano,sfikas2017exploiting} or multi-view \citep{su2015multi,wei2016dense,kalogerakis20173d} projections.

\paragraph*{Fixed operator methods.} Some methods use operators on surfaces to construct convolution-like operations. Surface Networks \cite{kostrikov2018surface} use discrete Laplacian and Dirac operators as edge weights in GNNs. \citet{yi2017syncspeccnn} define kernels in Laplacian eigenbases, including spectral parameterizations of dilated convolutional kernels and transformer networks.  \citet{qiao2020learning} use Laplacian spectral clustering to define neighborhoods for pooling.

\paragraph*{Learned operators.}

Some past methods learn relevant differential operators to a geometric learning task.  Closest to ours, \citet{wang2019learning} learn a parameterized sparse operator for geometry processing; see \S\ref{sec:operatorconstruction} for comparison of our operator to theirs.  Their layers simulate iterations of algorithms like conjugate gradients by applying their operator, limiting its receptive field to the number of layers.  \changed{In contrast, we explicitly perform eigendecomposition in our differentiable pipeline, allowing us to engineer the inductive bias inspired by the Hodge Laplacian.}  Similar discretizations are found in methods like \citep{eliasof2020diffgcn} for learning PDEs from data; this method uses algebraic multigrid to increase the receptive field. 

\paragraph*{Other.}

We mention a few other methods for learning from meshes that do not fall into the categories above.  \citet{xu2017directionally} present a pipeline that combines purely local and mesh-wise global features; \citet{feng2019meshnet} also propose extracting purely local features.  \citet{lim2018simple} apply recurrent neural networks (RNNs) to compute vertex features after unrolling local neighborhoods into prescribed spiral patterns.  Deep functional maps \citep{litany2017deep} largely rely on precomputed features for geometric information, although some recent efforts bring this correspondence method closer to end-to-end \citep{donati2020deep,sharma2020weakly}.

\paragraph*{Concurrent and unreviewed work.}  Machine learning is a fast-paced discipline, with new papers released daily.  Here, we acknowledge some ``late-breaking'' \changed{concurrent work}.

\citet{sharp2020diffusion} propose a ``learned diffusion layer'' in which features are diffused along a geometric domain via the isotropic heat equation with learned amount of diffusion; they include diffusion time as a learnable parameter.  Similarly to \citep{bruna2013spectral}, their diffusion is implemented in a fixed low-frequency Laplacian eigenbasis, computed during learning/inference.  Additional features incorporate anisotropy via inner products of spatial gradients. Unlike our work, they use a prescribed Laplacian operator.

Other methods include \citep{de2020gauge}, which proposes anisotropic gauge-invariant kernels using a message passing scheme built from parallel transport; \citep{lahav2020meshwalker}, an RNN-based approach employing random walks; \citep{schneider2020medmeshcnn}, which improves MeshCNN's memory efficiency and resilience to class imbalance for medical applications; \changed{and \citep{budninskiy2020laplacian}, which optimizes for a graph Laplacian parameterized by edge features}.

\section{Overview}

\begin{figure}
\includegraphics[trim=0 25 0 0,clip,width=\linewidth]{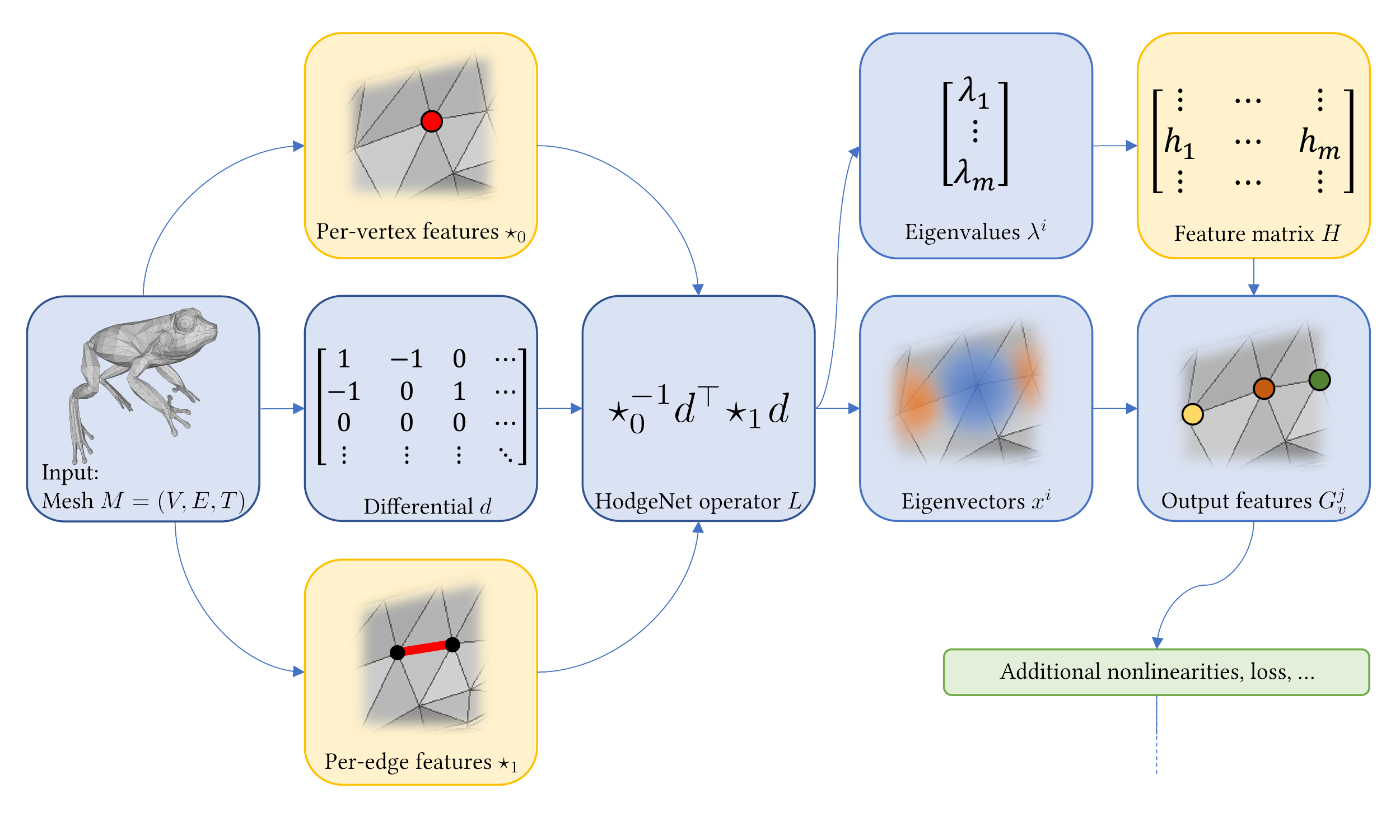}
\caption{Data flow in HodgeNet; yellow boxes contain learnable parameters.}\label{fig:data}
\end{figure}

Figure \ref{fig:data} gives an overview of our HodgeNet architecture for learning from triangle meshes.  The boxes highlighted in yellow have learnable parameters, while the remaining boxes are fixed computations.

Our goal is to learn an operator and associated spectral descriptor for a given learning-from-meshes task.  As with most methods, the learning stage uses stochastic gradient descent to optimize for model parameters, which are fixed during inference.

Our model inputs a triangle mesh $M=(V,E,T)$ and constructs three objects:
\begin{itemize}
\item a combinatorial \emph{differential} matrix $d\in\{-1,0,1\}^{|E|\times|V|}$,
\item a diagonal \emph{0-form Hodge star} matrix $\star_0\in\R^{|V|\times|V|}$, and
\item a diagonal \emph{1-form Hodge star} matrix $\star_1\in\R^{|E|\times|E|}$.
\end{itemize}
The matrix $d$ is a fixed function of $M$, while $\star_0,\star_1$ are learnable functions of local neighborhoods around mesh vertices.

HodgeNet then computes the $k$ eigenvectors $x^i$ of the semidefinite Laplacian-type matrix $L=\star_0^{-1}d^\top\star_1 d$ whose eigenvalues $\lambda^i$ are closest to zero.  Finally, per-vertex or per-mesh features are gathered from $\{(x^i,\lambda^i)\}$ using learnable formulas that generalize the form of popular spectral features like the heat kernel signature.

During training, we need to differentiate our loss function through the steps above.  Most of the operations above are simple nonlinearities that can be differentiated using standard backpropagation methods.  We show in \S\ref{sec:derivatives} how to obtain approximate derivatives of the eigenproblem efficiently.

\section{Operator Construction}\label{sec:operatorconstruction}

HodgeNet relies on a parameterized class of learnable operators whose entries are functions of local geometry.  The basis of our construction, designed to encapsulate operator constructions in spectral geometry, resembles that proposed by \citet{wang2019learning}, with the key difference that we expose per-edge and per-vertex features using diagonal Hodge star operators; this difference greatly simplifies our backpropagation procedure in \S\ref{sec:derivatives}. In \S\ref{sec:vectorial}, we also show how to generalize this construction for vectorial operators.

\subsection{Operator}

Given an oriented manifold mesh $M=(V,E,T)$ (optionally with boundary) with vertices $V\subset \R^3$, edges $E\subset V\times V$, and oriented triangles $T\subset V\times V\times V$, HodgeNet constructs a positive (semi)definite operator matrix $L\in\R^{|V|\times|V|}$ whose spectral structure will be used for a mesh-based learning task.

Inspired by the factorization of the Laplacian in discrete exterior calculus \citep{desbrun2005discrete}, we parameterize $L$ as a product:
\begin{equation}\label{eq:operator}
    L=\star_0^{-1} d^\top \star_1 d.
\end{equation}
Here, $d\in\{-1,0,1\}^{|E|\times|V|}$ is the differential operator given by
$$
d_{ev}=\left\{
\begin{array}{rl}
1 & \textrm{ if }v=v_2\\
-1 & \textrm{ if }v=v_1\\
0 & \textrm{ otherwise},
\end{array}
\right.
$$
where $e=(v_1,v_2)$ is an oriented edge.

While $d$ is determined by mesh topology, the diagonal Hodge star matrices $\star_0\in\R^{|V|\times|V|}$ and $\star_1\in\R^{|E|\times|E|}$ are learnable functions of local mesh geometry.  To construct $\star_0,\star_1$, we input $D$ per-vertex features $F\in\R^{|V|\times D}$. \changed{In our experiments, we use positions and normals as the per-vertex features, except when noted otherwise.}  We detail the construction of these operators $\star_0(F),\star_1(F)$ from $F$ below.

\paragraph{Per-vertex features $\star_0$.} Our construction of $\star_0\changed{(F)}$ imitates area weight computation for discrete Laplacians.  It takes place in two steps.  First, we compute \emph{per-triangle} features using a learnable function $f_\Phi:\R^{3D}\to\R$, where $\Phi$ contains the parameters of our model.  To ensure positive (semi)definiteness for $\star_0\changed{(F)}$ we square $f_\Phi$.  Finally, we gather features from triangles to vertices by summing and optionally adding a small constant $\epsilon$ \changed{(in practice, $\epsilon=10^{-4}$)} to improve conditioning of the eigensystem.  Overall, we can write our expression as follows:
\begin{equation}\label{eq:star0vv}
    (\star_0\changed{(F)})_{vv}=\epsilon+\sum_{t\sim v} f_\Phi(F_{v_1},F_{v_2},F_{v_3})^2
\end{equation}
where $t=(v_1,v_2,v_3)\in T$ is a triangle with vertices $v_1,v_2,v_3$ in counterclockwise order.  This sum over $t$ has a potentially different number of terms for each vertex, equal to the valence.

If $\epsilon=0$ and $f_\Phi^2$ measures triangle area scaled by $\nicefrac13$, then $\star_0$ becomes the barycentric area weights matrix often used in finite elements and discrete exterior calculus.  We give the details of our choice of functions $f_\Phi$ in \S\ref{sec:detailsandparameters}.  Squaring the inner part of \eqref{eq:star0vv} is one of many ways to make sure $(\star_0)_{vv}\geq0$ and could be replaced, e.g., by ReLU activation, but we found empirically that this simple expression led to the best performance.

\begingroup
\paragraph{Per-edge features $\star_1$.} The diagonal matrix $\star_1\changed{(F)}$ contains per-edge features on its diagonal.  Unlike \eqref{eq:star0vv}, to compute $\star_1\changed{(F)}$ we do not need to gather features from a variable-sized ring.  Instead, we learn a function $g_\Phi : \R^{4D} \to \R$ and, for an interior edge $e=(v_1,v_2)$, compute
\begin{equation}\label{eq:star1ee}
    (\star_1\changed{(F)})_{ee} = \epsilon + g_\Phi(F_{v_1},F_{v_2},F_{v_3},F_{v_4})^2,
\end{equation}
\setlength{\columnsep}{.1in}
\begin{wrapfigure}[5]{r}{.4\linewidth}\centering\vspace{-.3in}
\includegraphics[width=\linewidth]{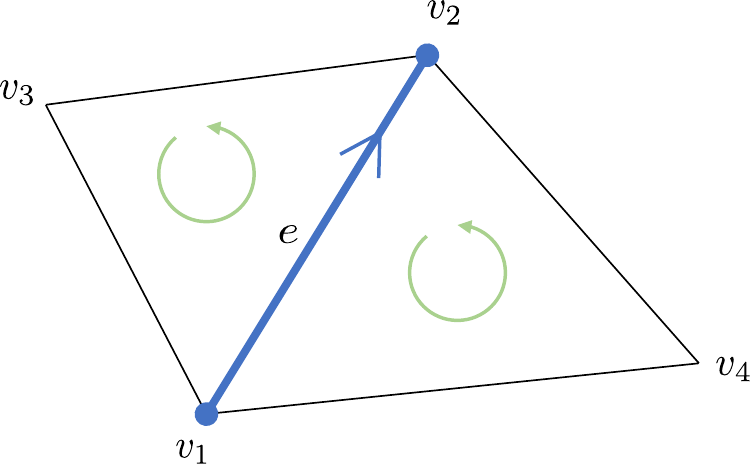}
\end{wrapfigure}
where $v_3$ and $v_4$ are \emph{opposite} the edge $e$ as shown to the right.  We order $v_3$ and $v_4$ so that $(v_1,v_2,v_3)$ and $(v_2,v_1,v_4)$ are all consistently oriented.  We learn a separate function $\bar g_\Phi(v_1,v_2,v_3)$ for boundary edges, since there is only one opposite \changed{vertex} in this case.

\endgroup

If $\epsilon=0$ and $g_\Phi^2$ gives the sum of interior angle cotangents at $v_3$ and $v_4$, then $\star_1$ will be the famous cotangent Laplacian matrix common in geometry processing.  While we have chosen to square the function $g$, thanks to conjugation by $d$ in \eqref{eq:operator} this is sufficient but not necessary for positive (semi)definiteness of $L$, and indeed this design choice prevents us from exactly reproducing the cotangent Laplacian in the presence of obtuse triangles.  Our architecture could easily be adjusted to allow for negative $(\star_1)_{ee}$ values and hence to reproduce the cotangent Laplacian operator, but the stability and ease of squaring $g_\Phi$ to ensure that $L$ has no negative eigenvalues outweighed this largely theoretical consideration.

\paragraph{Discussion.} Our parameterizations of $L$, $\star_0$, and $\star_1$ largely imitate the flow of information used to construct discrete Laplacian operators and related objects.  They are readily incorporated into geometry processing pipelines and have familiar sparsity patterns encountered in this discipline.

It is worth acknowledging a few design decisions intended to simplify our framework at the cost of mathematical structure:
\begin{itemize}
    \item Squaring $g_\Phi$ in \eqref{eq:star1ee} means we cannot reproduce the cotangent Laplacian operator for poorly-conditioned meshes with negative cotangent weights.
    \item We arbitrarily choose one of three possible cyclic orderings of the inputs to $f_\Phi$ in \eqref{eq:star0vv}.
    \item Similarly, we arbitrarily choose among two orderings of the inputs to $g_\Phi$ in \eqref{eq:star1ee}:  $(v_1,v_2,v_3,v_4)$ and $(v_2,v_1,v_4,v_3)$.
\end{itemize}
All three items above could be addressed at the cost of increasing the complexity of $f_\Phi,g_\Phi$, but building more general semidefiniteness conditions and/or order invariance did not bring practical benefit.

\subsection{Vectorial operators}\label{sec:vectorial}

$L$ discretizes operators that act on functions discretized using one value per vertex of a triangle mesh.  We also can discretize operators acting on \emph{vector-valued} functions with a value in $\R^k$ per vertex by adjustmenting our construction.  For example, for planar triangle meshes and $k=2$ we can reproduce the Killing operator described in \citep{solomon2011killing,claici2017isometry}; for $k=4$ we can mimic the Dirac operator used for shape analysis by \citet{liu2017dirac}.

To extend to vectorial operators, we use a $k|E|\times k|V|$ block version of $d$ whose blocks are given as follows:
$$
d_{ev}=\left\{
\begin{array}{ll}
I_{k\times k} & \textrm{ if }v=v_2\\
-I_{k\times k} & \textrm{ if }v=v_1\\
0 & \textrm{ otherwise},
\end{array}
\right.
$$
where $I_{k\times k}$ denotes the $k\times k$ identity matrix.

We generalize $f_\Phi:\R^{3D}\to\R^{k\times k}$ and $g_\Phi:\R^{4D}\to\R^{k\times k}$ to output $k\times k$ matrices.  Then, we compute $\star_0$ and $\star_1$ as block diagonal matrices whose elements are as follows:
\begin{align}
    (\star_0)_{vv}&=\epsilon I_{k\times k}+\sum_{t\sim v} f_\Phi(F_{v_1},F_{v_2},F_{v_3})^\top f_\Phi(F_{v_1},F_{v_2},F_{v_3})\\
    (\star_1)_{ee}&= \epsilon I_{k\times k} + g_\Phi(F_{v_1},F_{v_2},F_{v_3},F_{v_4})^\top g_\Phi(F_{v_1},F_{v_2},F_{v_3},F_{v_4}).
\end{align}
These definitions generalize our scalar construction for the case of $k=1$ and still lead to a semidefinite matrix $L=\star_0^{-1}d^\top \star_1 d\in\R^{k|V|\times k|V|}.$

\section{Differentiable Spectral Analysis}\label{sec:derivatives}

Now that we have determined the form of our operator $L$, we turn the task of using its low-order eigenvalues and eigenvectors for learning tasks.  The key challenge will be to differentiate through eigenvalue/eigenvector computation, a task we consider below.  While general eigencomputation is extremely difficult for learning, we show how our particular form \eqref{eq:operator} for $L$ facilitates backpropagation and reduces dependence on random-access computation.

Recall that a step of training requires evaluating the loss function $\L$ and its gradients with respect to the parameters $\Phi$ of the model.  The loss function is evaluated in a \emph{forward pass}, and the gradients are evaluated during \emph{backpropagation}.  We will perform both the forward and backward pass of our model on the CPU so as to take advantage of a sparse solver to \changed{compute} a set of eigenvalues/eigenvectors for the operator $L$ \changed{efficiently}.  While the computation of our features for our learning problem as well as the entirety of backpropogation could be efficiently computed on the GPU, our model has sufficiently few parameters that we find it unnecessary to transfer data between GPU and CPU.

\subsection{The HodgeNet \changed{Generalized} Eigenproblem}\label{sec:eigenproblem}

Our architecture outputs features built from eigenvectors of $L$ in \eqref{eq:operator}.  Recall that $L$---and, in particular, the Hodge stars $\star_0,\star_1$---is built from a matrix $F\in\R^{|V|\times D}$ of per-vertex features:
$$L=\star_0(F)^{-1} d^\top \star_1(F) d.$$
Hence, our features are built from eigenvectors $x^i\in\R^{k|V|}$ satisfying
\begin{equation}\label{eq:eigenproblem}
    Lx^i=\lambda^i x    ^i\iff d^\top\star_1 dx^i=\lambda^i\star_0x^i.
\end{equation}
By construction, $d^\top\star_1 d\succeq0$ and $\star_0\succeq0$, so $\lambda^i\geq0$.  By convention, we normalize eigenvectors to satisfy the condition $(x^i)^\top \star_0 x^j=\delta_{ij}$, possible thanks to symmetry of our operators.

To differentiate our per-vertex features, we need to differentiate the eigenvectors $x^i$ and eigenvalues $\lambda^i$ with respect to the parameters $\Phi$ of our learned functions $f_\Phi, g_\Phi$.  The expressions in \S\ref{sec:operatorconstruction} for $(\star_0)_{vv}(F)$ and $(\star_1)_{ee}(F)$ are readily differentiated.  Hence, for compatibility with the backpropagation algorithm for differentiation, we need to solve the following problem involving our loss function $\L$:  
\begin{center}
\textbf{Given the partial derivatives $\nicefrac{\partial\L}{\partial \lambda^i}$ and $\nicefrac{\partial\L}{\partial x_j^i}$ for all $i,j$, compute the partial derivatives $\nicefrac{\partial\L}{\partial(\star_0)_{vv}}$ and $\nicefrac{\partial\L}{\partial(\star_1)_{ee}}$ for all $v\in V,e\in E$.}
\end{center}
In words, given derivatives of the loss function with respect to the eigenvalues and eigenvectors of $L$, compute the derivatives of the loss function with respect to the Hodge stars.

In general, differentiating through eigenvalue problems is expensive.  Libraries like TensorFlow and PyTorch allow for differentiation of computing the \emph{full} spectrum of a matrix, but their implementations (1) cannot account for the sparsity structure of our mesh and (2) cannot target a few eigenvalues close to $0$, which are typically the meaningful eigenvalues to compute in geometry processing applications.   Solving the full eigenvalue problem is extremely expensive computationally, and storing a $k|V|\times k|V|$ matrix of eigenvectors is prohibitive.

Our pipeline addresses the issues above.  We rely on CPU-based sparse eigensolvers during the forward pass of our network, solving \eqref{eq:eigenproblem} only for a subset of eigenvalues.  This alleviates dependence on $k|V|\times k|V|$ dense matrices and instead only stores the $O(k|V|)$ nonzero entries.

\subsection{Derivative Formulas}

The vectorial operator $L$ operates on vectors in $\R^k$ per vertex on a mesh. Following \S\ref{sec:vectorial}, we will use $x^{i}_{v\ell}$ to refer to the $\ell$-th element ($\ell\in\{1,\ldots,k\}$) of entry $v$ ($v\in V$) of the $i$-th eigenvector of $L$.  
We use $\star_{0v\ell m}$ to refer to the element $(\ell,m)$ of the $k\times k$ block of $\star_0$ at vertex $v\in V$.
More generally, we will use subscripts to refer to matrix elements and superscripts to index over eigenvalues.

Define the following tensors:
\begin{align*}
    y_{e\ell}^i &:=\sum_{v\in V} d_{ev} x_{v\ell}^i\\
    M_{ij} &:=  \left\{
    \begin{array}{ll}
    (\lambda^i-\lambda^j)^{-1} & \textrm{ if }i\neq j\\
    0 & \textrm{ otherwise.}
    \end{array}
    \right.\\
    N_{ij} &:=\left\{
    \begin{array}{ll}
    \nicefrac{\lambda^i}{(\lambda^j-\lambda^i)} & \textrm{ if }i\neq j\\
    -\nicefrac{1}{2} & \textrm{ otherwise.}
    \end{array}
    \right.
\end{align*}
We compute $y$ during the forward pass as $dx$ and cache the result for use during backpropagation, since $d$ is a sparse matrix.

Our algorithm relies on the following proposition:
\begin{proposition}\label{prop:derivatives}
We can backpropagate derivatives of our loss function as follows:
{\allowdisplaybreaks
\begin{equation}\label{eq:grad}
\begin{array}{r@{\,}l}
    \displaystyle\frac{\partial \L}{\partial \star_{0w\ell m}} &= \displaystyle
    -\sum_i\frac{\partial \L}{\partial \lambda^i}\lambda^i x_{w\ell}^i x_{wm}^i
    +
    \sum_{ivnj} \frac{\partial \L}{\partial x^i_{vn}} N_{ij} x_{w\ell}^j x_{wm}^i x^j_{vn}\\
    \displaystyle\frac{\partial \L}{\partial \star_{1e\ell m}} 
    &\displaystyle=
    \sum_i\frac{\partial \L}{\partial \lambda^i}y_{e\ell}^i y_{em}^i
    +
    \sum_{ivnj} \frac{\partial \L}{\partial x^i_{vn}} M_{ij} x_{vn}^j y_{e\ell}^j y_{em}^i
    \end{array}
\end{equation}
}
Here, $i,j$ index over the eigenvectors of $L$; $\ell,n,m$ index over vector elements from $1$ to $k$; $v,w$ are vertices of the mesh; and $e$ is an edge.
\end{proposition}
\noindent We defer proof to the supplemental document, since it requires a fairly involved computation.  That said, this proposition is roughly an application of standard derivative-of-eigenvalue formulae to our operator $L$ in \eqref{eq:operator}, which benefits from the fact that our differentiable parameters are in \emph{diagonal} matrices $\star_0,\star_1$.

The expressions in \eqref{eq:grad} may appear complicated, but in reality they are efficiently computable.  We have eliminated all sparse matrices and inverses from these formulas, which are readily implemented using a one-line call to Einstein summation functions in deep learning toolkits (e.g., \texttt{einsum} in PyTorch).

\subsection{Derivative Approximation}\label{sec:derivapprox}

Here we briefly address one challenge using Proposition \ref{prop:derivatives} to differentiate HodgeNet.  Recall from \S\ref{sec:eigenproblem} that we compute an incomplete set of eigenvectors of $L$, far fewer than the largest possible number.  This choice is reasonable for constructing a loss function, which will only depend on this low-order eigenstructure.  However, \eqref{eq:grad} requires \emph{all} eigenvectors of $L$ to evaluate the sums over $j$. 

We use a simple strategy to address this issue.  During the forward pass we compute and cache more eigenvalues/eigenvectors than are needed to evaluate $\L$; in practice, we use $2\times$ (see \S\ref{sec:ablation} for validation).  Then, in backpropagation we truncate the sums over $j$ in \eqref{eq:grad} to include only these terms.  A straightforward argument reveals that the resulting gradient approximation still yields a descent direction for $\L$.

The first term in each sum is computable from exclusively the partial set of eigenvalues, implying we can exactly differentiate $\L$ with respect to the eigenvalues $\lambda^i$; our approximation is only relevant to the eigenvectors.

\section{From Eigenvectors to Features}\label{sec:eigtofeature}


Recall that our broad task is to design a learnable mapping from meshes to task-specific features.  So far, we have designed a learnable operator from mesh geometry and provided a means of differentiating through its eigenvectors/eigenvalues.  It is tempting to use the eigenvectors as per-vertex features, but this is not a suitable choice:  The choice of sign $\pm x_i$ for each eigenvector is arbitrary.

We return to geometry processing for inspiration.  Classical shape descriptors built from operator eigenfunctions circumvent the sign issue by \emph{squaring} the Laplacian eigenfunctions pointwise.  For instance, the heat kernel signature \citep{sun2009concise}, wave kernel signature \citep{aubry2011wave}, and general learned kernels \cite{litman2013learning} take the form
$$\sum_i f(\bar\lambda^i) \psi^i(p)^2,$$
where $\bar\lambda^i$ is the $i$-th eigenvalue and $\psi^i$ is the $i$-ith eigenvector of the Laplacian.  The fact that $\psi^i$ is squared alleviates sign dependence.  Similarly, for eigenfunctions of the vectorial \emph{vector Laplacian}, sign-agnostic features can be computed from the outer product of the pointwise vector and itself $\psi^i(p)\psi^i(p)^\top\in\R^{k\times k}$ (see, e.g., \cite[eq.\ (3.13)]{singer2012vector}).

Generalizing the hand-designed features above, we construct a sign-agnostic learnable per-vertex feature as follows.  Take $m$ to be the number of eigenvectors of $L$ we will use to compute features, and take $n$ to be the number of output features.  We learn a function $h_\Phi:\R\to\R^n$ and construct a matrix $H\in\R^{m\times n}$ whose columns are $h(\lambda^i)$ for $i\in\{1,\ldots,m\}$.  Then, for $j\in\{1,\ldots,n\}$, the $j$-th output feature at vertex $v\in V$, notated $G_v^j$, is given by:
$$G_v^j :=\sum_{i} H_{ij}\cdot (x^i_v)(x^i_v)^\top,$$
where $x^i_v$ denotes the $i$-th eigenvector of $L$ evaluated at vertex $v$ as a $k\times 1$ column vector. We omit the 0 eigenvalue corresponding to the constant eigenfunction. We give our form for $h_\Phi$ in \S\ref{sec:detailsandparameters}.

Having computed per-vertex features $G_v$, we optionally follow a standard max pooling approach to obtain per-face features
$$G_f = \max_{v \sim f} G_v,$$ or per-mesh features $$G_M = \max_{v \in V} G_v$$ depending on the learning task at hand.  We map these features to the desired output dimensions $d$ using a learned function $o_\Phi : \R^n \to \R^d$.

\section{Additional Details and Parameters}\label{sec:detailsandparameters}

We model each of $f_\Phi$, $g_\Phi$, $h_\Phi$, $o_\Phi$ as an MLP with batch normalization and Leaky ReLU nonlinearity \cite{maasrectifier} before each hidden layer. $f_\Phi$ and $g_\Phi$ each consist of four hidden layers, each of of size 32; $h_\Phi$ consists of four hidden layers, each of size $n$; and $o_\Phi$ consists of two hidden layers, each of size 32, except for the classification task, where the layers have 64 units. In all our experiments, we set vector dimensionality $k=4$, output feature size $n=32$, and number of eigenpairs used $m=32$. We use an additional 32 eigenpairs for improved derivative approximation, as described in \S\ref{sec:derivapprox}.

We train our network using the optimizer AdamW \cite{loshchilov2018decoupled} with a batch size of 16 and learning rate of 0.0001. We use gradient clipping with maximum norm of 1.0 to stabilize training. We implement our pipeline in PyTorch, using SciPy \texttt{eigsh} with ARPACK for solving our sparse eigenproblem and \texttt{libigl} for mesh processing. We train our models on 128 2.5 GHz CPUs.

\section{Experiments}

We demonstrate the efficacy of our method on several shape analysis tasks and provide experiments justifying some of our parameter and design choices.  \changed{We also compare to state-of-the-art methods developed for learning on meshes. Other geometric deep learning approaches tend to use GNNs, ignoring the structure and relying on multiple layers to aggregate global data, whereas our method uses spectral geometry to infer global information from local features.}

\subsection{Mesh Segmentation}

We train our network for the task of mesh segmentation on four datasets---the Human Body dataset \cite{maron2017convolutional} and the vase, chair, and alien categories of the Shape COSEG dataset \cite{wang2012active}---optimizing the standard cross entropy loss. We use the same version of the Human Body dataset as in \cite{hanocka2019meshcnn, milano2020primal}, which is downsampled to 2000 faces per mesh. We evaluate on the test set of the Human Body dataset, and generate a random 85\%-15\% train-test split for each Shape COSEG category, as in \cite{hanocka2019meshcnn, milano2020primal}. We train for 100 epochs (about 3 hours), randomly decimating each input mesh to a resolution of 1000-2000 faces and randomly applying anisotropic scaling of up to 5\% in each dimension. We then fine-tune by training for 100 more epochs without decimation or scaling. In the case of the Human Body dataset, where meshes are not canonically rotated, we also apply random rotations as data augmentation to the training set. We center each mesh about the vertex center of mass and rescale to fit inside the unit sphere.

\begin{table}[h]
    \centering
    \begin{tabular}{ccc}
         \toprule
         Method & \# Parameters & Accuracy \\
         \midrule
         Ours & 31,720 & 85.03\% \\
         PD-MeshNet \cite{milano2020primal} & 173,728 & 85.61\%\\
         MeshCNN \cite{hanocka2019meshcnn} & 2,279,720 & 85.39\%\\
         \bottomrule
    \end{tabular}
    \caption{Segmentation accuracy on the Human Body test set.}
    \label{tab:human_seg}
\end{table}

\begin{table}[h]
    \centering
    \begin{tabular}{cc}
         \toprule
         Method & Accuracy \\
         \midrule
         Ours &  86.48\% \\
         PD-MeshNet \cite{milano2020primal} & 86.45\%\\
         \bottomrule
    \end{tabular}
    \caption{Area-weighted segmentation accuracy on Human Body test set.}
    \label{tab:human_seg_weighted}
\end{table}

\begin{figure}[h]
    \centering
    \includegraphics[width=\linewidth]{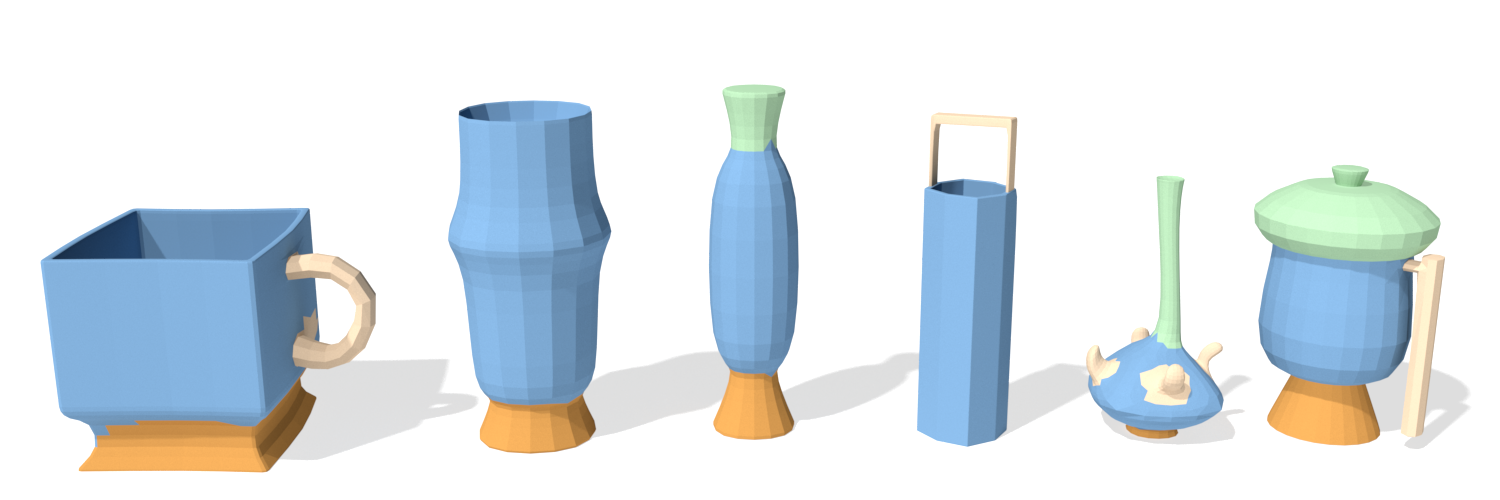}
    \vspace{0.1in}
    \includegraphics[width=\linewidth]{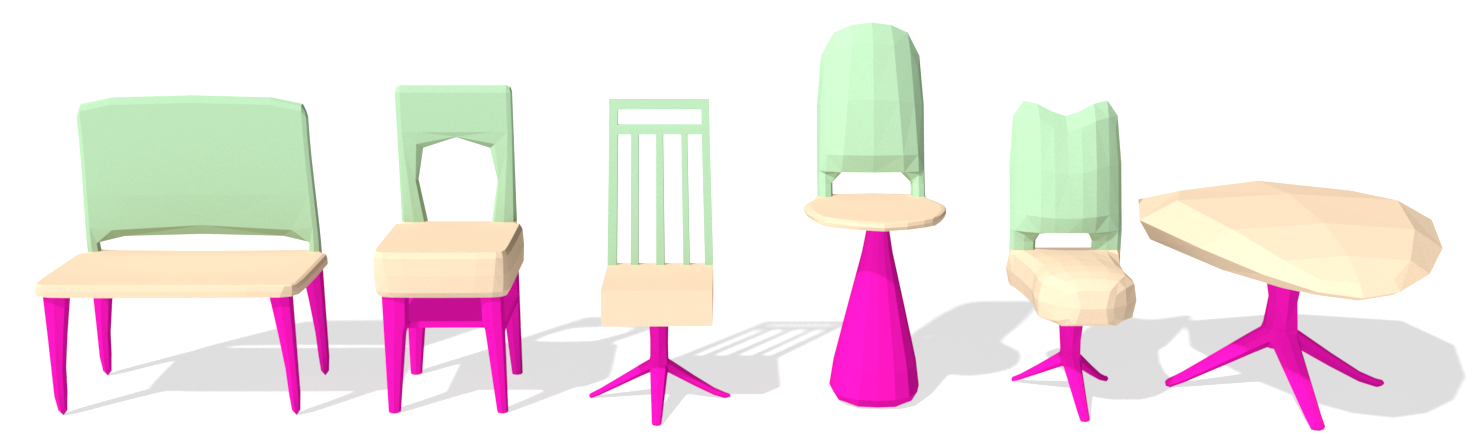}
    \vspace{0.1in}
    \includegraphics[width=\linewidth]{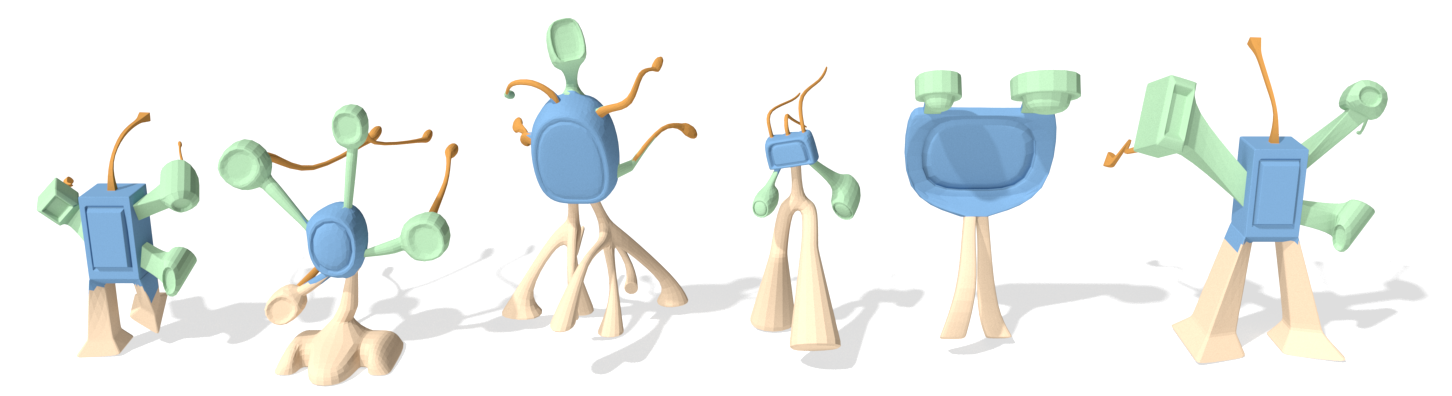}
    \caption{Segmentation results on the Shape COSEG dataset. Meshes shown above are randomly selected from the test set for each category.}
    \label{fig:coseg}
\end{figure}

\begin{figure*}[h]
    \centering
    \includegraphics[width=\linewidth]{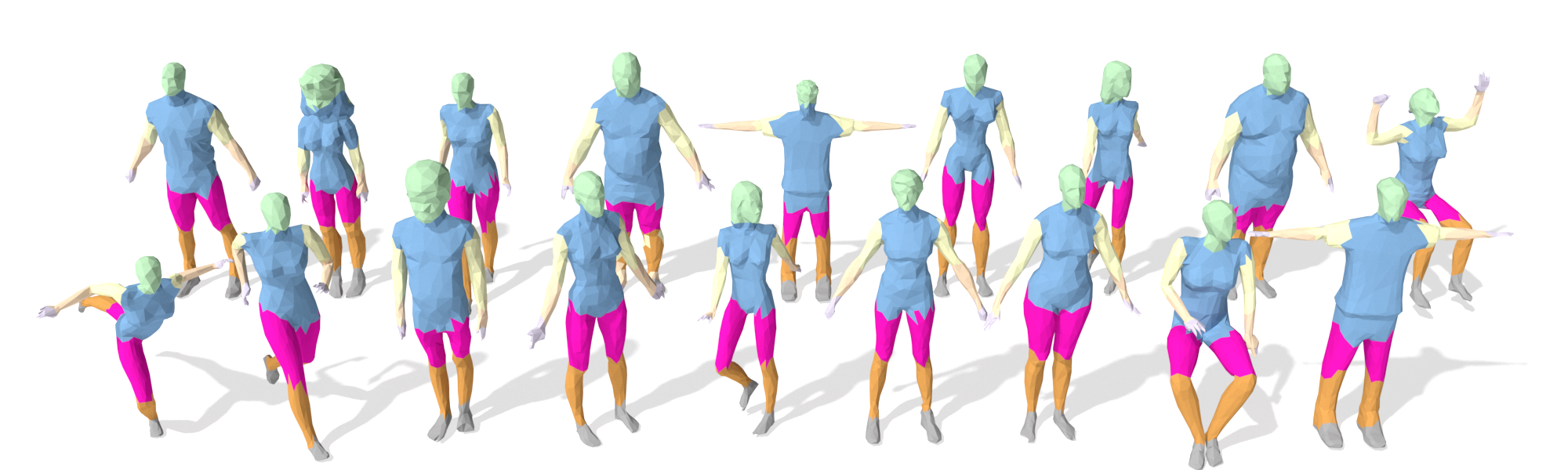}
    \caption{Mesh segmentation results on the Human Body test set.}
    \label{fig:human_seg}
\end{figure*}

\begin{table}[h]
    \centering
    \begin{tabular}{cccc}
         \toprule
         Method & Vases & Chairs & Aliens \\
         \midrule
         Ours &  90.30\% & 95.68\%  & 96.03\%   \\
         PD-MeshNet \cite{milano2020primal} & 95.36\% & 97.23\% & 98.18\% \\
         MeshCNN \cite{hanocka2019meshcnn} & 92.36\% & 92.99\% & 96.26\% \\
         \bottomrule
    \end{tabular}
    \caption{Test segmentation accuracy on Shape COSEG.}
    \label{tab:coseg}
\end{table}

\begin{table}[h]
    \centering
    \begin{tabular}{cccc}
         \toprule
         Method & Vases & Chairs & Aliens \\
         \midrule
         Ours &  94.38\% & 99.22\%  & 97.97\%   \\
         PD-MeshNet \cite{milano2020primal} & 97.49\% & 97.86\% & 98.66\% \\
         \bottomrule
    \end{tabular}
    \caption{Area-weighted test segmentation accuracy on Shape COSEG.}
    \vspace{-0.2in}
    \label{tab:coseg_weighted}
\end{table}

We report segmentation accuracies in Tables~\ref{tab:human_seg} and~\ref{tab:coseg} \changed{and area-weighted segmentation accuracies in Tables~\ref{tab:human_seg_weighted} and \ref{tab:coseg_weighted}. For fair comparison, as in \cite{milano2020primal}, we report accuracies based on ``hard" ground-truth segmentation face labels for MeshCNN \cite{hanocka2019meshcnn} rather than ``soft" edge labels; see  \cite[Supplementary Material, Section H]{milano2020primal} for details regarding the segmentation metrics.} Our method obtains results comparable to state-of-the-art for each dataset while requiring significantly fewer learnable parameters. We also show our learned segmentations on the entire Human Body test set in Figure~\ref{fig:human_seg} and on a sampling of the Shape COSEG test sets in Figure~\ref{fig:coseg}.

\subsection{High-Resolution Mesh Segmentation}

\begingroup
\setlength{\columnsep}{0.1in}
\begin{wraptable}[11]{r}{.4\linewidth}\centering\vspace{-0.15in}
    \centering
    \begin{tabular}{cc}
         \toprule
         Split \# & Accuracy \\
         \midrule
         1 & 90.57\%  \\
         2 & 86.90\% \\
         3 & 90.02\% \\
         4 & 89.07\% \\ 
         5 & 90.15\% \\
         \bottomrule
    \end{tabular}
    \caption{Mesh segmentation accuracies for five random test splits of the full-resolution MIT animation dataset. }
    \label{tab:humans_highres}
\end{wraptable}
In contrast to earlier works, our method is capable of training on dense, non-decimated mesh data. We demonstrate this by training a segmentation model on the MIT animation dataset \cite{vlasic2008articulated}, where each mesh consists of 20,000 faces. We pre-initialize our model with the segmentation model trained on the Human Body dataset above and train for an additional 30 epochs, approximately 4 hours. \changed{The pre-initialization allows us to save training time by avoiding training a model from scratch: our model trained on low-resolution mesh data is able to capture some triangulation-invariant features, making this transfer learning possible.} We train on five randomly sampled 95\%-5\% train-test splits and achieve a mean accuracy of 89.34\%. We report segmentation accuracies for each split in Table~\ref{tab:humans_highres} and render Split 1 in Figure~\ref{fig:humans_highres} and Split 2 in Figure~\ref{fig:humans_highres2}.

\begin{figure}[h]
    \centering
    \includegraphics[width=\linewidth]{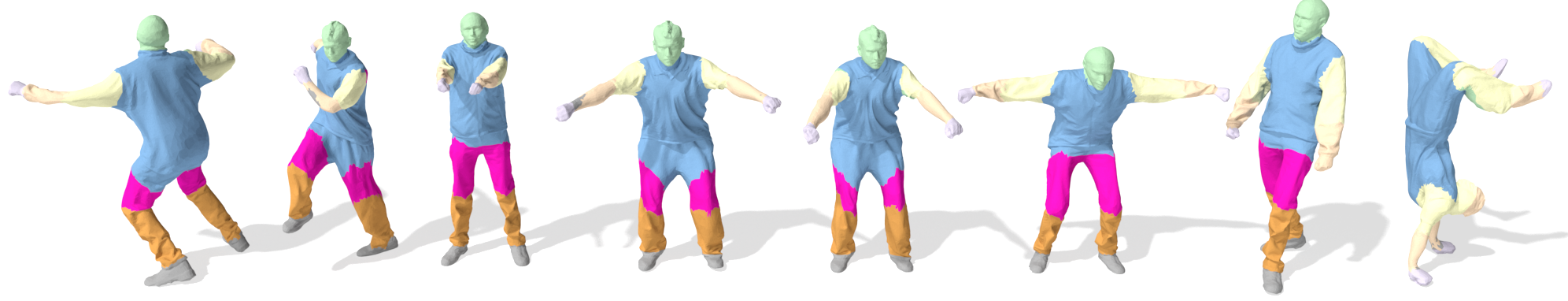}
    \caption{Test set mesh segmentations of Split 2  of the full-resolution MIT animation dataset.}
    \label{fig:humans_highres2}
\end{figure}
\endgroup

\subsection{Mesh Classification}

We evaluate our method on mesh classification on the downsampled version of the SHREC dataset \cite{3DOR:3DOR11:079-088}, as in \cite{hanocka2019meshcnn, milano2020primal}, optimizing the standard cross entropy loss. We report our results on two different splits of the dataset---\emph{Split 10}, where each of the 30 shape categories is randomly split into 10 test and 10 training examples, and \emph{Split 16}, where each category is split into 4 test and 16 training examples---in Table~\ref{tab:shrec}.  We train for 100 epochs using decimation to 400-500 faces, anisotropic scaling, and random rotations as data augmentation and then fine-tune for another 100 epochs for \emph{Split 16} and 200 epochs for \emph{Split 10}.

\begin{table}[h]
    \centering
    \begin{tabular}{ccc}
         \toprule
         Method & Split 16 & Split 10  \\
         \midrule
         Ours & 99.17\% & 94.67\% \\
         PD-MeshNet \cite{milano2020primal} & 99.7\% & 99.1\% \\
         MeshCNN \cite{hanocka2019meshcnn} & 98.6\% & 91.0\% \\
         \bottomrule
    \end{tabular}
    \caption{Classification accuracy on the SHREC test set.}
    \label{tab:shrec}
\end{table}

\subsection{Dihedral Angle Prediction}

As a stress test, we demonstrate that our method is capable of learning an operator that is sensitive to extrinsic geometry. To this end, we propose a synthetic dataset for dihedral angle regression. Previous methods that rely on computing a Laplacian would necessarily fail at this task, as they are only aware of intrinsic structure.

\begingroup

\setlength{\columnsep}{0.1in}
\begin{wrapfigure}[5]{r}{.3\linewidth}\centering\vspace{-.15in}
\includegraphics[width=\linewidth]{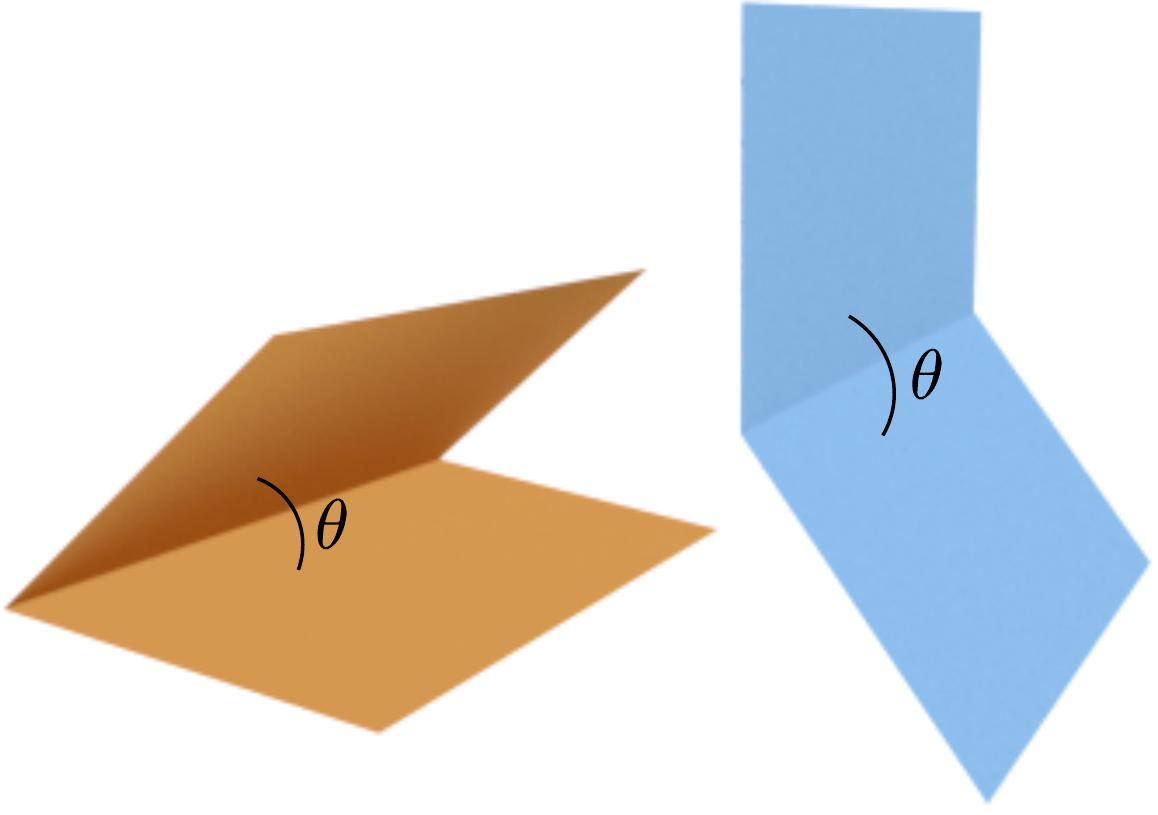}
\end{wrapfigure}
We take a regular mesh of a flat square consisting of 100 faces and crease it down the center at a random angle $\theta \in [0, 2\pi]$, as shown. Our network learns a two-dimensional vector per mesh, and we optimize cosine distance to the ground truth $\theta$. We use the same hyperparameters as for the other experiments with a batch size of 32. For this experiment, we only use vertex positions as the input features---we do not provide normals. After training for just 15 minutes, we are able to predict the angle with an average error of $0.17\degree$.

\endgroup

\subsection{Ablation}\label{sec:ablation}

We perform an ablation study to justify some the design and parameter choices in our architecture. In Table~\ref{tab:ablation}, we report test accuracy on the Shape COSEG vases dataset after 100 epochs of training (without fine-tuning). The accuracy degrades when we do not provide normals as part of the input mesh features, when we do not cache any additional eigenpairs for improved derivative approximation, when we reduce the vector dimensionality $k$, when we reduce the learned feature size $n$, or when we use fewer eigenpairs $m$ for feature computation.

\begin{table}[h]
    \centering
    \begin{tabular}{cc}
         \toprule
         Model & Accuracy \\
         \midrule
         full & 87.78\% \\
         no normals & 86.26\% \\
         no additional eig. & 87.44\% \\
         $k=2$ & 79.02\% \\
         $m=16$ & 86.34\% \\
         $n=8$ & 87.08\% \\
         \bottomrule
    \end{tabular}
    \caption{Ablation study of our parameter choices on segmentation of the Shape COSEG vases dataset.}
    \label{tab:ablation}
\end{table}



\section{Discussion and Conclusion}

HodgeNet has many features that make it an attractive alternative for learning from meshes.  During inference, its structure resembles that of most spectral geometry processing algorithms:  construct a useful operator, and compute features from its spectrum.   Our model is lightweight in the sense that the learnable functions act only on local neighborhoods, yet our model has a global receptive field thanks to the eigenvector computation.  It has relatively few parameters and can be evaluated efficiently on the CPU.

This exploratory work suggests many avenues for future research.  The most obvious next step is to extend our model to tetrahedral meshes for volumetric problems; we do not anticipate any major issues with this extension.  We also can use our method's connection to DEC to make learnable versions of other discrete differential operators, e.g.\ ones acting on $k$-forms for $k\geq1$, and we can consider other applications of learning on meshes like generative modeling.

Our work also reveals some insight into other learning problems.  Our architecture could easily be applied to graphs rather than triangle meshes, essentially by mildly changing the parameterization of $\star_1$ and taking $\star_0$ to be the identity matrix; we hence anticipate that there may be some applications to network analysis and other graph learning problems.  Our lightweight differentiation strategy for eigenvectors may also prove useful in other contexts demanding eigenstructure of large matrices.

From the broadest perspective, our work demonstrates one of many potential applications of differentiable sparse linear algebra.  While our derivative approximations and specially-formulated operator provide one way to circumvent development of a general framework for combining deep learning and linear algebra, a framework coupling sparse linear algebra to deep learning toolkits would enable a vast set of modeling choice and applications currently hamstrung by available architectures.

\begin{acks}
The MIT Geometric Data Processing group acknowledges the generous support of Army Research Office grant W911NF2010168, of Air Force Office of Scientific Research award FA9550-19-1-031, of National Science Foundation grant IIS-1838071, from the CSAIL Systems that Learn program, from the MIT–IBM Watson AI Laboratory, from the Toyota--CSAIL Joint Research Center, from a gift from Adobe Systems, from an MIT.nano Immersion Lab/NCSOFT Gaming Program seed grant, and from the Skoltech--MIT Next Generation Program. This work was also supported by the National Science Foundation Graduate Research Fellowship under Grant No. 1122374.
\end{acks}

\bibliographystyle{ACM-Reference-Format}
\bibliography{hodgenet}



\section*{Supplementary material}

*Supplemental Material for*
# HodgeNet: Learning Spectral Geometry on Triangle Meshes

In this document, our goal is to derive derivative formulas for the eigenvalue/eigenvector problem associated to a *block* diagonal Hodge star operator, for vector-valued functions on meshes. We will do so using tensor-style notation, which is directly compatible with implementation in deep learning libraries.

## 1 SETUP

As input, assume we have a matrix $d \in \mathbb{R}^{|E| \times |V|}$ that subtracts per-vertex values along mesh edges. Our goal is to learn two operators $\star_0 \in \mathbb{R}^{|V| \times k \times k}$ and $\star_1 \in \mathbb{R}^{|E| \times k \times k}$, where $k$ is the dimensionality of the vectorial data.

Suppose $x \in \mathbb{R}^{|V| \times k}$ is a $k$-valued function on our domain. Then, the weak form of our Laplace-type operator acting on $x$ can be written $x \in \mathbb{R}^{|V| \times k} \mapsto \mathfrak{L}[x] \in \mathbb{R}^{|V| \times k}$ where

$$\mathfrak{L}[x]_{v\ell} = \sum_{emw} d_{ev} \star_{1e\ell m} d_{ew} x_{wm}. \tag{1}$$

Similarly, we can define a mass operator as

$$\mathfrak{M}[x]_{v\ell} = \sum_{m} \star_{0v\ell m} x_{vm}. \tag{2}$$

Throughout this document, we will assume the symmetry properties

$$\star_{0v\ell m} = \star_{0vm\ell} \quad \text{and} \quad \star_{1e\ell m} = \star_{1em\ell}. \tag{3}$$

In a forward pass, we compute a set of eigenvectors $x^i$ and eigenvalues $\lambda^i$ satisfying

$$\mathfrak{L}[x^i] = \lambda^i \mathfrak{M}[x^i]. \tag{4}$$

Or equivalently, plugging in (1) and (2), for all $i, v, \ell$ we have

$$\sum_{emw} d_{ev} \star_{1e\ell m} d_{ew} x^i_{wm} = \lambda^i \sum_{m} \star_{0v\ell m} x^i_{vm} \tag{5}$$

Because our operators are self-adjoint, we have the following orthogonality relationship:

$$\sum_{v\ell} x^j_{v\ell} \mathfrak{M}[x^i]_{v\ell} = \delta_{ij}. \tag{6}$$

Or equivalently, plugging in (2),

$$\sum_{v\ell m} x^j_{v\ell} \star_{0v\ell m} x^i_{vm} = \delta_{ij}. \tag{7}$$

Suppose our learning problem as an loss function $\mathcal{L}$. During the process of back propagation, we end up with derivatives

$$\frac{\partial \mathcal{L}}{\partial x^i_{vj}} \quad \text{and} \quad \frac{\partial \mathcal{L}}{\partial \lambda^i}. \tag{8}$$

In this document, we show how to obtain

$$\frac{\partial \mathcal{L}}{\partial \star_{0v\ell m}} \quad \text{and} \quad \frac{\partial \mathcal{L}}{\partial \star_{1e\ell m}}. \tag{9}$$

## 2 USEFUL IDENTITIES

We start by deriving some general formulas. We will use prime to denote differentiation with respect to a scalar we will determine later; note $\mathfrak{L}$ and $\mathfrak{M}$ are functions of our unknowns in the learning problem so we need to be careful taking derivatives.

We'll first differentiate (5). For all $i, v, \ell$ we have

$$0 = \left( \sum_{emw} d_{ev} \star_{1e\ell m} d_{ew} x^i_{wm} - \lambda^i \sum_{m} \star_{0v\ell m} x^i_{vm} \right)'$$

$$= \sum_{emw} \left( d_{ev} \star'_{1e\ell m} d_{ew} x^i_{wm} + d_{ev} \star_{1e\ell m} d_{ew} (x^i_{wm})' \right) - \sum_{m} \left( (\lambda^i)' \star_{0v\ell m} x^i_{vm} + \lambda^i \star'_{0v\ell m} x^i_{vm} + \lambda^i \star_{0v\ell m} (x^i_{vm})' \right)$$

We multiply both sides by $x_{v\ell}^j$ and sum over $v, \ell$ to obtain the following:

$$\begin{aligned}
0 &= \sum_{emwv\ell} x_{v\ell}^j \left( d_{ev} \star'_{1e\ell m} d_{ew} x_{wm}^i + d_{ev} \star_{1e\ell m} d_{ew} (x_{wm}^i)' \right) \\
&\quad - \sum_{mv\ell} x_{v\ell}^j \left( (\lambda^i)' \star_{0v\ell m} x_{vm}^i + \lambda^i \star'_{0v\ell m} x_{vm}^i + \lambda^i \star_{0v\ell m} (x_{vm}^i)' \right) \\
&= \sum_{emwv\ell} x_{v\ell}^j \left( d_{ev} \star'_{1e\ell m} d_{ew} x_{wm}^i + d_{ev} \star_{1e\ell m} d_{ew} (x_{wm}^i)' \right) \\
&\quad - (\lambda^i)' \delta_{ij} - \lambda^i \sum_{mv\ell} x_{v\ell}^j \left( \star'_{0v\ell m} x_{vm}^i + \star_{0v\ell m} (x_{vm}^i)' \right) \text{ by (7)} \\
&= \sum_{emwv\ell} x_{v\ell}^j d_{ev} \star'_{1e\ell m} d_{ew} x_{wm}^i + \sum_{v\ell}(x_{v\ell}^i)' \sum_{emw} d_{ev} \star_{1e\ell m} d_{ew} x_{wm}^j \\
&\quad - (\lambda^i)' \delta_{ij} - \lambda^i \sum_{mv\ell} x_{v\ell}^j \left( \star'_{0v\ell m} x_{vm}^i + \star_{0v\ell m} (x_{vm}^i)' \right) \text{ reindexing/reshuffling the second term} \\
&= \sum_{emwv\ell} x_{v\ell}^j d_{ev} \star'_{1e\ell m} d_{ew} x_{wm}^i + \lambda^j \sum_{v\ell m}(x_{v\ell}^i)' \star_{0v\ell m} x_{vm}^j \\
&\quad - (\lambda^i)' \delta_{ij} - \lambda^i \sum_{mv\ell} x_{v\ell}^j \left( \star'_{0v\ell m} x_{vm}^i + \star_{0v\ell m} (x_{vm}^i)' \right) \text{ by (5)} \\
&= \sum_{emwv\ell} x_{v\ell}^j d_{ev} \star'_{1e\ell m} d_{ew} x_{wm}^i + (\lambda^j - \lambda^i) \sum_{v\ell m}(x_{v\ell}^i)' \star_{0v\ell m} x_{vm}^j \\
&\quad - (\lambda^i)' \delta_{ij} - \lambda^i \sum_{mv\ell} x_{v\ell}^j \star'_{0v\ell m} x_{vm}^i \text{ after reindexing the last term} \\
&= \sum_{em\ell} y_{e\ell}^j \star'_{1e\ell m} y_{em}^i + (\lambda^j - \lambda^i) \sum_{v\ell m}(x_{v\ell}^i)' \star_{0v\ell m} x_{vm}^j - (\lambda^i)' \delta_{ij} - \lambda^i \sum_{mv\ell} x_{v\ell}^j \star'_{0v\ell m} x_{vm}^i,
\end{aligned} \quad (10)$$

where we define

$$y_{e\ell}^j := \sum_v d_{ev} x_{v\ell}^j. \quad (11)$$

This leads us to a useful identity:

$$(\lambda^i)' \delta_{ij} + (\lambda^i - \lambda^j)\langle (x^i)', x^j \rangle_0 = \sum_{ew} y_{e\ell}^j \star'_{1e\ell m} y_{em}^i - \lambda^i \sum_{v\ell m} x_{v\ell}^j \star'_{0v\ell m} x_{vm}^i, \quad (12)$$

where $\langle \cdot, \cdot \rangle_0$ denotes an inner product with respect to $\star_0$.

For one more convenient formula, we can differentiate (7) in the case $i = j$:

$$\begin{aligned}
0 &= \sum_{v\ell m} (x_{v\ell}^i \star_{0v\ell m} x_{vm}^i)' \\
&= \sum_{v\ell m} (x_{v\ell}^i \star'_{0v\ell m} x_{vm}^i + 2 x_{v\ell}^i \star_{0v\ell m} (x_{vm}^i)') \text{ by symmetry of } \star_0 \\
\implies \langle x^i, (x^i)' \rangle_0 &= -\frac{1}{2} \sum_{v\ell m} x_{v\ell}^i \star'_{0v\ell m} x_{vm}^i \quad (13)
\end{aligned}$$

## 3 DERIVATIVES WRT $\star_1$

We start with the $\star_1$ operator. By the chain rule, we know

$$\frac{\partial \mathcal{L}}{\partial \star_{1e\ell m}} = \underbrace{\sum_i \frac{\partial \mathcal{L}}{\partial \lambda^i} \frac{\partial \lambda^i}{\partial \star_{1e\ell m}}}_{\text{term 1}} + \underbrace{\sum_{ivn} \frac{\partial \mathcal{L}}{\partial x_{vn}^i} \frac{\partial x_{vn}^i}{\partial \star_{1e\ell m}}}_{\text{term 2}} \quad (14)$$

Consider (12) with $i = j$ and differentiating with respect to $\star_{1e\ell m}$. We find:

$$\frac{\partial \lambda^i}{\partial \star_{1e\ell m}} = y_{e\ell}^i y_{em}^i \quad (15)$$

Differentiating the eigenvectors is more difficult. We use a projection formula:

$$\frac{\partial x^i_{vn}}{\partial \star_{1e\ell m}} = \sum_j \left\langle \frac{\partial x^i}{\partial \star_{1e\ell m}}, x^j \right\rangle_0 x^j_{vn} \text{ by writing in the } x \text{ basis}$$

$$= \sum_{j \neq i} \left\langle \frac{\partial x^i}{\partial \star_{1e\ell m}}, x^j \right\rangle_0 x^j_{vn} \text{ by (13)}$$

$$= \sum_j M_{ij} x^j_{vn} y^j_{e\ell} y^i_{em} \text{ by (12) with } i \neq j, \tag{16}$$

where

$$M_{ij} = \begin{cases} (\lambda_i - \lambda_j)^{-1} & \text{if } i \neq j \\ 0 & \text{otherwise.} \end{cases} \tag{17}$$

Plugging these expressions into (14),

$$\frac{\partial \mathcal{L}}{\partial \star_{1e\ell m}} = \sum_i \frac{\partial \mathcal{L}}{\partial \lambda^i} y^i_{e\ell} y^i_{em} + \sum_{ivnj} \frac{\partial \mathcal{L}}{\partial x^i_{vn}} M_{ij} x^j_{vn} y^j_{e\ell} y^i_{em} \tag{18}$$

## 4 DERIVATIVES WRT $\star_0$

Now, we consider the $\star_0$ operator. By the chain rule, we have

$$\frac{\partial \mathcal{L}}{\partial \star_{0w\ell m}} = \underbrace{\sum_i \frac{\partial \mathcal{L}}{\partial \lambda^i} \frac{\partial \lambda^i}{\partial \star_{0w\ell m}}}_{\text{term 1}} + \underbrace{\sum_{ivn} \frac{\partial \mathcal{L}}{\partial x^i_{vn}} \frac{\partial x^i_{vn}}{\partial \star_{0w\ell m}}}_{\text{term 2}}. \tag{19}$$

Consider (12) with $i = j$ and differentiating with respect to $\star_{0w\ell m}$. We find:

$$\frac{\partial \lambda^i}{\partial \star_{0w\ell m}} = -\lambda^i x^i_{w\ell} x^i_{wm} \tag{20}$$

Differentiating the eigenvectors is again more difficult. We use a projection formula:

$$\frac{\partial x^i_{vn}}{\partial \star_{0w\ell m}} = \sum_j \left\langle \frac{\partial x^i}{\partial \star_{0w\ell m}}, x^j \right\rangle_0 x^j_{vn} \text{ by writing in the } x \text{ basis}$$

$$= -\frac{1}{2} x^i_{w\ell} x^i_{wm} x^i_{vn} + \sum_{j \neq i} \left\langle \frac{\partial x^i}{\partial \star_{0w\ell m}}, x^j \right\rangle_0 x^j_{vn} \text{ by (13)}$$

$$= -\frac{1}{2} x^i_{w\ell} x^i_{wm} x^i_{vn} - \sum_{j \neq i} M_{ij} \lambda^i x^j_{w\ell} x^i_{wm} x^j_{vn} \text{ by (12) with } i \neq j$$

$$= \sum_j N_{ij} x^j_{w\ell} x^i_{wm} x^j_{vn} \tag{21}$$

where

$$N_{ij} = \begin{cases} -\frac{\lambda^i}{\lambda^i - \lambda^j} & \text{if } i \neq j \\ -\frac{1}{2} & \text{otherwise.} \end{cases} \tag{22}$$

Plugging these expressions into (23),

$$\frac{\partial \mathcal{L}}{\partial \star_{0w\ell m}} = -\sum_i \frac{\partial \mathcal{L}}{\partial \lambda^i} \lambda^i x^i_{w\ell} x^i_{wm} + \sum_{ivnj} \frac{\partial \mathcal{L}}{\partial x^i_{vn}} N_{ij} x^j_{w\ell} x^i_{wm} x^j_{vn} \tag{23}$$



\end{document}